\documentstyle[preprint,aps]{revtex}
\tightenlines
\begin{document}
\title{"Cold Melting" of Invar Alloys}
\author{M. Molotskii and V. Fleurov}
\address{
School of Physics and Astronomy, Beverly and Raymond Sackler
Faculty of Exact Sciences.\\ Tel Aviv University, Tel Aviv 69978,
Israel}
\date{\today}
\maketitle
\begin{abstract}
An anomalously strong volume magnetostriction in Invars may lead
to a situation when at low temperatures the dislocation free
energy becomes negative and a multiple generation of dislocations
becomes possible. This generation induces a first order phase
transition from the FCC crystalline to an amorphous state, and may
be called "cold melting". The possibility of the cold melting in
Invars is connected with the fact that the exchange energy
contribution into the dislocation self energy in Invars is
strongly enhanced, as compared to conventional ferromagnetics, due
to anomalously strong volume magnetostriction. The possible
candidate, where this effect can be observed, is a FePt disordered
Invar alloy in which the volume magnetostriction is especially
large.
\end{abstract}

\section{Introduction}
Among various approaches to the theory of crystal melting (see
review \cite{b87}) the dislocation model (see
\cite{m52,kw65,n78,ew79,ktt90,chhr91,bps00} and references
therein) occupies the leading position, describing the relevant
physical phenomena in the most adequate way. According to the
simplest version of the model \cite{kw65}, one should consider the
change of the crystal free energy $\Delta F=W-T\Delta S$, caused
by a single dislocation, calculated per its unit length. Here $W$
is the internal energy change, and $\Delta S$ is the corresponding
entropy change. At elevated temperatures the free energy $\Delta
F$ may become negative, so that a spontaneous multiple generation
of dislocations becomes thermodynamically favorable. This
generation destroys the long range order of the crystal and leads
to its amorphization and melting. A reasonable assumption is that
both the energy $W$ and the entropy $\Delta S$ are related only to
the elastic strains induced by the dislocation and, hence, they
hardly depend on temperature. It means that the melting
temperature can be determined by the equation
\begin{equation}\label{melt1}
  T^{hot}_m=\beta W_{el}a.
\end{equation}
with $a$ being the interatomic spacing. Here $\beta$ is a
proportionality coefficient, which is $\beta = 1/(a\Delta S)$ in
the simplest case \cite{kw65}. Generally, the coefficient $\beta$
may have a more complicated shape, since it should additionally
incorporates such effects, as interactions between the
dislocations, entropy due to various dislocation configurations
and so on (see discussion in \cite{ew79} and in the most recent
paper \cite{bps00}). The notation $W_{el}$ in equation
(\ref{melt1}) emphasizes connection of the dislocation self energy
to the elastic strains.

When considering ferromagnetic crystals, whose internal energy
incorporates also the exchange interaction energy, one may think
about an additional contribution due to the variation of this
exchange interaction in the field of the elastic strains around
the dislocations. In typical ferromagnetics, say, Fe or Ni, the
exchange interaction contribution makes at best several tenths of
one percent of the elastic energy and can be disregarded. However,
a very strong volume magnetostriction typical of the Invar alloys,
as demonstrated, e.g., in \cite{w90}, may result in an enormous
enhancement of the dislocation exchange self energy and become
comparable with the elastic one. Our recent paper \cite{mf00}
discusses the influence of this enhancement on the interaction
between dislocations and solute atoms in Invar alloys with the aim
to explain some features of plastic deformation of Invars.

The exchange self energy of a dislocation is negative and its
absolute value grows in the ferromagnetic phase with decreasing
temperature. It will be demonstrated in this paper that at low
enough temperatures ($T<T_C$, $T_C$ is the Curie temperature) the
total dislocation energy, which is now the sum of the elastic and
exchange contributions, decreases significantly with decreasing
temperature, and in some Invar alloys may even change its sign. It
means that the criterion (\ref{melt1}) can be met not only at
elevated temperatures but also at a low temperature. It may lead
to a situation when spontaneous multiple generation of
dislocations and crystal amorphization may become
thermodynamically favorable below a certain critical temperature.

This phenomenon may be called "cold melting". We plan to discuss
here the conditions of the cold melting and find the temperature
below which it becomes possible. We have no intentions to address
here in detail the problem of what is the state of the crystal
below the transition temperature and what the order of the
transition is. We may only mention that we do not currently see
reasons why the analysis of, say, Edwards and Warner \cite{ew79}
is not applicable in our case as well. They have demonstrated for
the usual hot melting that the dislocation mechanism results in a
first order phase transition. We expect that a similar first order
transition may lead to an amorphized glassy-like state at low
temperatures.

In order to find the temperature of the cold melting we need to
calculate the dislocation self energy. First, the elastic energy
of a unit length of an edge dislocation is determined by the
equation \cite{hl82}
\begin{equation}\label{en1}
  W_{el}=\frac{\mu b^2}{4\pi(1-\nu)}ln\frac{R}{r_0}.
\end{equation}
Here $b$ is the magnitude of the dislocation Burgers vector, $\mu$
is the shear modulus, $\nu$ is the Poisson coefficient. When
calculating the energy (\ref{en1}), the integral of the
dislocation induced strain field is cut both at small distances
$r_0$, of the order of the interatomic spacing, and at large
distances $R$, of the order of the average distance between the
dislocations.

The volume magnetostriction results in a variation of the exchange
energy of a ferromagnetic in the strain field in the vicinity of a
dislocation. In order to calculate the corresponding energy change
we consider the density of the exchange energy of the
ferromagnetic in the molecular field approximation (see,
e.g.,\cite{v74})
\begin{equation}\label{en2}
{\cal  W}_{ex}=-\frac{\omega M^2}{2}
\end{equation}
where $M$ is the local magnetization, and
\begin{equation}\label{coef1}
\omega=\frac{3k_BT_C}{np^2_{eff}\mu_B^2}
\end{equation}
is the constant of the molecular field. Here $n$ is the atomic
density, $p_{eff}$ is the effective number of the Bohr magnetons
$\mu_B$ per atom. In principle, one should also consider a term
proportional to $(\nabla M)^2$ in the energy (\ref{en2}). Its
contribution to the exchange energy is, however, two orders of
magnitude smaller than the leading term in (\ref{en2}) and, hence,
it has been neglected.

The hydrostatic pressure $p$ created by an edge dislocation causes
a local change of the magnetization
\begin{equation}\label{magn1}
  M=\overline M(1+\alpha p)
\end{equation}
where $\overline M$ is the spontaneous magnetization of the
ferromagnetic in the absence of the dislocation, $\alpha$ is a
proportionality coefficient known empirically. It is usually
rather small but may take anomalously high values in Invar alloys
\cite{hm81,ks60}.

The hydrostatic pressure in the vicinity of an edge dislocation is
\cite{hl82}
\begin{equation}\label{pres1}
p(\rho,\theta) = - \frac{\mu b}{3\pi} \frac{1+\nu}{1-\nu}
\frac{\sin\theta}{\rho}
\end{equation}
with $\rho$ and $\theta$ being the cylindrical coordinates.
Substituting (\ref{magn1}) and (\ref{pres1}) into the energy
(\ref{en2}) and integrating, using the same cuts as when
calculating the elastic energy, one finds the exchange self energy
of a dislocation per its unit length
\begin{equation}\label{en3}
W_{ex}= - \frac{\omega\overline M^2\alpha^2b^2\mu^2}{18\pi}
\left(\frac{1+\nu}{1-\nu}\right)^2\ln\frac{R}{r_0}.
\end{equation}
Now adding the exchange energy (\ref{en3}) to the elastic one
(\ref{en1}), the total dislocation self energy per its unit length
takes the form
\begin{equation}\label{en4}
  W = W_{el} + W_{ex} = f(T) W_{el}
\end{equation}
where
\begin{equation}\label{func1}
  f(T) = 1 - \frac{\omega\overline M^2(T)\alpha^2(T)E(T)}{9}
\frac{1+\nu(T)}{1-\nu(T)}.
\end{equation}
$E=2\mu(1+\nu)$ is the Young modulus. It is emphasized in equation
(\ref{func1}) that material parameters in Invars are temperature
dependent.

A similar analysis can be carried out with respect to the
interaction between the dislocations. It leads to the conclusion
that equation (\ref{en4}) is, in fact, more general. A similar
equation with the same factor (\ref{func1}) holds also for the
total internal energy of the system of dislocations which includes
both the internal self energies of the individual dislocation and
the interactions between them.

The second term in equation (\ref{func1}) reflects the
contribution of the dislocation exchange energy. It is usually
very small, about 10$^{-3}$, in conventional ferromagnetics and,
hence, may be neglected. In Invars, however, the situation changes
dramatically. It is connected with the fact that the coefficient
$\alpha$, connected to the volume magnetostriction, may be several
tens times larger than, say, in Fe or Ni \cite{ks60}. In the
ferromagnetic phase this second term grows with lowering
temperature and may become comparable to one, so that the function
$f(T)$ may become very small or even change its sign.

Now we discuss how the exchange contribution to the dislocation
energy influences the melting criterion (\ref{melt1}) for the
temperature at which the spontaneous generation of dislocations
becomes possible. Direct analysis shows that now we should take
the same condition (\ref{melt1}) but substitute there the total
self energy of the dislocation (\ref{en4}) instead of the elastic
one. As for the entropy due the dislocation induced changes in the
magnetic system, it can be still neglected. Then the melting
temperature can be found as a solution of the equation
\begin{equation}\label{melt2}
T_m = \beta f(T_m)W_{el}a.
\end{equation}

This equation may have more than one solutions. One solution can
be found at high temperatures and corresponds to the conventional
hot melting temperature $T_m^{(hot)}$. Usually the melting
temperature exceeds the Curie temperature where the local
magnetization disappears, $\overline M=0$, hence, $f(T)=1$. Then
equation (\ref{melt2}) coincides with (\ref{melt1}) and leads to
the standard description of the hot melting. As for the cold
melting temperature, it is connected with hot melting temperature
by the condition
\begin{equation}\label{melt3}
  T_m^{(cold)}=T_m^{(hot)}f(T_m^{(cold)})
\end{equation}

Equation (\ref{melt3}) can be solved when using the low
temperature values of the material parameters and Curie-Weiss
equation
\begin{equation}\label{magn2}
\overline M(T) = M_0\sqrt{1-\frac{T}{T_C}}
\end{equation}
for the spontaneous magnetization. Here $M_0$ is the spontaneous
magnetization at zero temperature. Then
\begin{equation}\label{melt4}
T_m^{(cold)} = T_C\frac{\gamma - 1}{\gamma - \displaystyle
\frac{T_C}{T_m^{(hot)}}}
\end{equation}
where
\begin{equation}\label{gamma}
\gamma = \frac{\omega M^2_0\alpha^2E}{9}\frac{1+\nu}{1-\nu}
\end{equation}

In the known ferromagnetics $T_C < T_m^{(hot)}$, hence, the
temperature of the cold melting may be positive only if $\gamma >
1$. Although rather rough approximations have been used (see
discussion of Invar properties in \cite{w90}), when deriving
equation (\ref{melt4}), the condition $\gamma >1 $ is in fact more
general and reflects just the fact that the energy necessary for
creation of a dislocation becomes negative. This condition can be
used as a good indicator when looking for Invar alloys in which
cold melting may be observed.

It can be worth considering Fe-Pt Invar alloys, with a content
close to Fe$_{0.72}$Pt$_{0.28}$, which are characterized by high
values of $\alpha$. The volume magnetostriction becomes especially
large in Fe$_{0.72}$Pt$_{0.28}$ alloys with disordered
distribution of Fe and Pt atoms over the lattice sites. It may
reach the value $\alpha =-2.4\times 10^{-11}$(dyn/cm$^2$)$^{-1}$
at room temperatures \cite{hm81}. As for a detailed information on
the temperature dependence of $\alpha$ in these alloys,
unfortunately, it is not available. Nevertheless, we shall use
this value in the estimates to be done below.

This alloy is characterized by the following parameters:
$T_C=371$K, $p_{eff}=2.13$ \cite{ssn76}, and $n=7.6\times
10^{22}$cm$^3$. Then equation (\ref{coef1}) leads to
$\omega=5190$. At $M_0=np_{eff}\mu_B=1.5$kG, $\nu = 0.3$, and
$E=1.2 \times 10^{12}$ dyn/cm$^2$ one gets $\gamma = 1.66$.
Therefore, we see that the condition $\gamma > 1$ is holds very
well.

Then knowing the temperature of the hot melting,
$T_m^{(hot)}=1812$K, for this alloy one could have found the
temperature $T_m^{(cold)}=168$K for the cold melting which follows
from equation (\ref{melt4}). However, we prefer doing the same
calculation by using the experimental data on the temperature
dependence of the magnetization $\bar M(T)/M_0$ \cite{ssn76}, and
elastic constants \cite{h74} temperature. The resulting
temperature is even higher $T_m^{(cold)}=298$K and is rather close
to the room temperature. This also provides a justification for
using the room temperature value of the parameter $\alpha$ in this
estimate.

Therefore, the prediction of our model is that at temperatures
below $T_m^{(cold)}=298$K a spontaneous multiple generation of
dislocations Fe-Pt Invar alloys in may become possible. It is
interesting to note that an intensive generation of dislocations
is really observed in disordered Fe-Pt Invar alloys in the
temperature range from 300 to 77K in several experiments (see,
e.g., \cite{mof88} and references therein). Our estimate for the
temperature of the cold melting lies close to the upper limit of
this range. However, the experimental observed phenomenon is
rather complicated. This generation goes hand in hand with the
martensite transitions from FCC to BCC phase, also observed in the
same temperature range, and the two effects can be hardly
separated one from another.

There are reasons to believe that they are really tightly
connected. On one hand, dislocations may be generated as a result
of accumulation and discharge of elastic strains in the vicinity
of the boundaries between the martensite and austenite (host)
phases in the process of the martensite growth \cite{w96}. A
multiple generation of dislocations is observed only in disordered
alloys with larger values of the parameter $\alpha$ which favors
its connection with the mechanism of the cold melting discussed
above. As for ordered alloys with a smaller value of $\alpha$ a
dislocation generation is not observed in them.

On the other hand, we believe that the martensite transitions,
which are much more intensive in disordered Fe-Pt alloys
\cite{mof88}, may be themselves induced by the dislocation
generation in the process of the cold melting. It is known that
appearance of strain-induced martensites is strongly facilitated
by defects consisting of properly arranged dislocations which play
the role of active centers in the formation of the martensite
\cite{w96}. We assume that the dislocation generation leads to
creation of such centers, and these centers induce martensite
transitions.

A detailed experimental and theoretical study of the properties of
Invar alloys at temperatures close and below the cold melting
temperature present a special interest. One may expect highly
unusual plastic properties of Invars in this region. As an example
of what one may expect, we mention that at $T < T_m^{(cold)}$
dislocations of the same sign start attracting rather than
repelling each other, meaning that the theory of the deformation
hardening should be completely revisited.

\end{document}